\documentclass[aps,prx,superscriptaddress,twocolumn,showpacs, longbibliography]{revtex4-1}

\usepackage{amssymb}
\usepackage{amsmath}
\usepackage{amsthm}
\usepackage{graphicx}
\usepackage{amsfonts}
\usepackage{color}
\usepackage{times}
\usepackage{natbib}
\usepackage{comment}
\usepackage{soul}
\usepackage{hyperref}
\hypersetup{
        colorlinks=true,
}

\newcommand{\bra}[1]{\langle #1|}
\newcommand{\ket}[1]{|#1\rangle}

\newcommand{\pr}[1]{|#1\rangle\langle #1|}

\newcommand{\bea}{\begin{eqnarray}}
\newcommand{\eea}{\end{eqnarray}}
\newcommand{\myskip}[1]{}
\newcommand{\comments}[1]{}

\DeclareMathOperator{\tr}{Tr}
\newcommand{\proj}[1]{}
\newcommand{\iv}{\mathbf{i}}
\newcommand{\ivb}{\bar{\mathbf{i}}}

\DeclareMathOperator{\Tr}{Tr}

\begin{document}

\title{Extractable Work from Correlations}

\author{Mart\'i~Perarnau-Llobet$^{*}$}
\affiliation{ICFO-Institut de Ciencies Fotoniques, Mediterranean Technology Park, 08860 Castelldefels (Barcelona), Spain}

\author{Karen~V.~Hovhannisyan$^{*}$}
\affiliation{ICFO-Institut de Ciencies Fotoniques, Mediterranean Technology Park, 08860 Castelldefels (Barcelona), Spain}

\author{Marcus~Huber}
\affiliation{ICFO-Institut de Ciencies Fotoniques, Mediterranean
Technology Park, 08860 Castelldefels (Barcelona), Spain}
\affiliation{Departament de F\'{i}sica, Universitat Aut\`{o}noma
de Barcelona, E-08193 Bellaterra, Spain}

\author{Paul~Skrzypczyk}
\affiliation{ICFO-Institut de Ciencies Fotoniques, Mediterranean
Technology Park, 08860 Castelldefels (Barcelona), Spain}
\affiliation{H. H. Wills Physics Laboratory, University of Bristol, Tyndall Avenue, Bristol, BS8 1TL, United Kingdom}

\author{Nicolas~Brunner}
\affiliation{D\'epartement de Physique Th\'eorique, Universit\'e de Gen\`eve, 1211 Gen\`eve, Switzerland}

\author{Antonio~Ac\'in}
\affiliation{ICFO-Institut de Ciencies Fotoniques, Mediterranean
Technology Park, 08860 Castelldefels (Barcelona), Spain}
\affiliation{ICREA-Instituci\'o Catalana de Recerca i Estudis Avan\c cats, Lluis Companys 23, 08010 Barcelona, Spain\\
{\small $^*$(These authors equally contributed to this work.)}}

\begin{abstract}

Work and quantum correlations are two fundamental resources in thermodynamics and quantum information theory. In this work we study how to use correlations among quantum systems to optimally store work. We analyse this question for isolated quantum ensembles, where the work can be naturally divided into two contributions: a local contribution from each system, and a global contribution originating from correlations among systems. We focus on the latter and consider quantum systems which are locally thermal, thus from which any extractable work can only come from correlations. We compute the maximum extractable work for general entangled states, separable states, and states with fixed entropy. Our results show that while entanglement gives an advantage for small quantum ensembles, this gain vanishes for a large number of systems.

\end{abstract}

\maketitle

\section{Introduction}

Traditional, macroscopic thermodynamics is strikingly robust to the underlying mechanics: its three laws hold true while switching from classical to quantum mechanics \cite{ll5}. On the other hand, one would hope for the opposite, since thermodynamics is intimately connected to information theory \cite{koji}, and quantum phenomena, such as entanglement, have a drastic effect on the latter, irrespective of the scale \cite{mikeike}.

Recently much attention has been dedicated to the problem of understanding thermodynamics of small quantum systems. This has led notably to the development of a resource theoretical formulation of quantum thermodynamics \cite{spekkens,oppi,ljr} and, in a more practical vein, to the study of quantum thermal machines \cite{palao,noah,leko,ahmg,Lutz,luis,lutz,correa,pksk}. The role and significance of quantum effects 
to thermodynamics is still to be fully understood, although progress has recently been achieved \cite{alicki,karen,nicolas,lutz,Ueda,correa,pksk,funo,felo,rodi}.

A problem of particular importance in quantum thermodynamics is to understand which quantum states allow for the storage and extraction of work from quantum systems \cite{pusz,lenard}. Such states are called non-passive, while states from which no work can be extracted are referred to as passive. Remarkably, the latter have the property of activation: when considered as a whole, several copies of passive states can become non-passive. The only states lacking this property are the thermal (also referred to as completely passive) states \cite{pusz,lenard}. 

The situation changes when considering ensembles that can also be correlated. There, even a collection of locally thermal states can be non-passive \cite{parrondo,vaikuntanathan,ahepl}. The main goal of the present work is to understand how to optimally make use of correlations among quantum systems for work storage. Specifically, we consider a quantum ensemble composed of $n$ subsystems (particles or modes). Each subsystem is assumed to be in a thermal state, at the same temperature $T$. The total system, however, is correlated, because otherwise its state would also be thermal hence passive. This is in fact the natural scenario to study the role of correlations for work storage, as they become the only source of non-passivity.

First, we show that if no restriction on the global state is made, then it is possible to store in the system the maximal amount of work compatible with the requirement that the reduced states are thermal. In other words, at the end of the protocol, the system is left in the ground state and, thus, all energy has been extracted. Notably this is possible thanks to quantum entanglement. It is then natural to ask if the same amount of work can be stored using a separable, or even a purely classical state diagonal in the product energy eigenbasis, that is, with no coherences among different energy levels. We will see that, although the amount of work that can be stored in unentangled states is strictly smaller than the amount that can be stored in entangled states for any finite $n$, the gain decreases with the size of the system and in the thermodynamic limit ($n \rightarrow \infty$) purely classical states already become optimal. In fact, quantum resources offer a significant advantage only for small $n$, while neither entanglement nor energy coherences are needed for optimal work storage in the thermodynamic limit. We also consider additional natural constraints on the global state, such as limiting the entropy or requiring the decohered (classical) version of the state to be thermal, and investigate the role of quantum coherence and entanglement in these cases. 

Finally, we show that our results are also applicable in the scenario where the system has an access to a thermal bath. There the connection between work extraction and correlations have been studied before \cite{Oppenheim,AlickiHorodecki,Zurek,MMVedral,Jevtic,Lutz,Ueda,dahlsten,Vincent,braga,levtoff} Given access to global operations on the subsystems, the extractable work is proportional to the mutual information \cite{Oppenheim,Jevtic}. That is, only the strength of the correlations is relevant, and not the type (i.e. quantum or classical). Here, in contrast, we show that when the bath (a macroscopic object) is not available and one has only a few subsystems, quantum correlations do provide a sizeable advantage. This brings new insights in the quantum-to-classical transition in thermodynamics.

\section{Framework}

We consider an isolated quantum system which consists of $n$ $d$-level subsystems. 
The local Hamiltonian $h = \sum_a E_a \ket{a}\bra{a}$ is taken to be the same for each subsystem and, without loss of generality, it is assumed that the ground state energy is zero. We consider the situation where there is no interaction Hamiltonian between the subsystems, such that the total Hamiltonian $H$ is simply the sum of the individual local Hamiltonians $H = \sum_i h_i$. 

The class of operations that we consider is the class of \emph{cyclic Hamiltonian processes}, i.e. we can apply any time dependent interaction $V(t)$ between the $n$ subsystems for a time $\tau$, such that $V(t)$ is non-vanishing only when $0 \leq t \leq \tau$. The corresponding evolution can be described by a unitary operator $U(\tau)= \overrightarrow\exp\left(-i\int_0^\tau d t\left(H + V(t)\right)\right)$, where $\overrightarrow\exp$ denotes the time-ordered exponential. By varying over all $V(t)$ we can generate any unitary operator $U = U(\tau)$ and therefore this class of operations can alternatively been seen as the ability to apply any global unitary on the system.

The task we are interested in is work extraction via a cyclic Hamiltonian process. Since the system is taken to be isolated, there are no other systems to exchange energy with, therefore the extracted work is the change in average energy of the system under such a process \cite{commentwork}. More precisely, we define the extracted work $W$ as
\begin{equation}\label{e:work}
W = \Tr\left(\rho H \right)- \Tr\left(U\rho U^\dagger H\right) .
\end{equation}
Within this framework, it is well known that work can be extracted from a system if and only if the system is \emph{non-passive}, where a passive system with Hamiltonian $H = \sum_{\alpha} \mathcal{E}_\alpha \ket{\alpha}\bra{\alpha}$ ($\mathcal{E}_\alpha\leq\mathcal{E}_{\alpha+1}$) 
is the one whose state is of the form
\begin{align}
\rho^{\mathrm{passive}} = \sum_{\alpha} p_\alpha \pr{\alpha} 
\quad\text{with} \quad p_{\alpha+1}\leq p_\alpha.
\end{align}
In other words, a system is passive if and only if its state is diagonal in the energy eigenbasis and has eigenvalues non-increasing with respect to energy. Now it easily follows that, given a non-passive state $\rho$, the extracted work (\ref{e:work}) is maximized by \cite{armen}:
\begin{equation}
W_{\rm max}=\rm{Tr} (\rho H)-\rm{Tr} (\rho^{\mathrm{passive}} H)
\label{Wmax}
\end{equation}
where $\rho$ and $\rho^{\mathrm{passive}}$ have the same spectrum and therefore there exists a unitary operator taking the former to the latter. Equation (\ref{Wmax}) defines the energy that can be potentially extracted from the state via cyclic hamiltonian (unitary) processes. This quantity will be the main focus of this article, and we will refer to it as \emph{extractable work, stored work} or \emph{work content} (the term {\it ergotropy} is also used in the literature \cite{armen}).

Importantly, we see that passivity is a global property of a system, and thus this raises interesting possibilities when considering a system comprised of a number of subsystems, as we do here. Indeed, global operations are capable of extracting more work than local ones, as a state can be locally passive but globally not. Such an enhancing may have two origins: activation or correlations between subsytems. Activation occurs when $(\rho^{\mathrm{passive}})^{\otimes k}$ becomes a non-passive state for some $k$. Interestingly, thermal states are the only passive states that do not allow for activation, as any number of copies of thermal states is also thermal \cite{pusz,lenard}. On the other hand, states that are locally passive but have a non-product structure (i.e., they are correlated) also offer the possibility for work extraction. An extreme case, which is the focus of this article, is a set of correlated locally thermal states, as in such a case the global contribution uniquely comes from correlations. Our goal, in fact, is to understand how correlations allow for work extraction in systems that are locally completely passive \cite{commentsp}.

We will therefore focus on the subset of all possible states of the system, comprised by \emph{locally thermal} states, that is all $\rho$ such that the reduced state of subsystem $i$ satisfies
\begin{equation}\label{e:locally thermal}
\rho_i = \Tr_{\overline{i}}\rho = \tau_\beta
\end{equation}
for all $i$, where $\Tr_{\overline{i}}$ denotes the partial trace over all subsystems except subsystem $i$. Here $\tau_\beta$ is the thermal state of the subsystem at (a fixed but arbitrary) inverse temperature $\beta = 1/T$,
\begin{equation}\label{tb}
\tau_{\beta}=\frac{1}{\mathcal{Z}}e^{-\beta h},
\end{equation}
where $\mathcal{Z} = \tr e^{-\beta h}$ is the partition function.

Now, if $\rho$ is locally thermal \eqref{e:locally thermal}, and since $H$ is a sum of local Hamiltonians, the first term of the right hand side of \eqref{Wmax} is fixed and is given by $\Tr\left(\rho H \right) = n E_\beta$, where $E_\beta = \Tr(\tau_\beta h)$ is the average energy of the local thermal state. Note also that given our convention that the ground state has zero energy, the second term of the right hand side of \eqref{Wmax}, that is, the final average energy, is always nonnegative. This implies that the extractable work is upper bounded by
\begin{equation}
\label{wmax}
W_{\rm max}\leq n E_\beta .
\end{equation}
This bound is attainable if and only if the final state is the ground state, denoted by $ \ket{0}^{\otimes n}$. 

Apart from understanding how to exploit the general correlations to store work in the system, we will also study the particular role of entanglement and energy coherences in these processes. We consider three natural sets of correlated states:
(i) arbitrary states, thus including entangled ones, (ii) separable
states and a subset of them: (iii) states diagonal in the product energy
eigenbasis. We will study work extraction for these three different sets of correlated quantum states.




Before proceeding further, we will end by noting that in the present context our quantity of interest is the \textit{average} extractable work. This allows us to obtain precise and quantitative results about the relation between work and quantum correlations in the initial state. The question of how to obtain similar results, for example about the full work probability distribution, in general remains a difficult open problem. Essentially, at the moment there is no framework allowing to obtain the full work distribution function of the process without destroying the initial coherences (and entanglement) of the state (see \cite{armenII} for a discussion on how to extent fluctuation theorems for coherent states).  







\section{Extractable work from correlations} \label{sec3}

We first show that within the above framework quantum correlations are capable of making all the energy in the system available for extraction in the form of work, as they allow saturating the bound (\ref{wmax}). As mentioned above, it can only be saturated if and only if $U\rho U^\dagger$ is the ground state. Now observe that the state
\begin{equation}
\label{e:StateEnt}
\ket{\phi}=\frac{1}{\sqrt{\mathcal{Z}}}\sum_{a=0}^{d-1} e^{-\frac{\beta E_a}{2}} \ket{a}^{\otimes n}.
\end{equation}
is locally thermal, i.e., such that $\Tr_{\overline{i}}\ket{\phi}\bra{\phi} = \tau_\beta$ for all $i$. Moreover, since it is pure, there exists a unitary matrix $U$ such that $U\ket{\phi} = \ket{0}^{\otimes n}$. Thus all the energy $nE_\beta$ can be extracted from state $\ket{\phi}$ and $W_{\rm max}=nE_\beta$.

However, it is clear that the state \eqref{e:StateEnt} is entangled. Hence it is natural to ask whether the amount of extractable work would change if we restrict ourselves to separable, or even classical  states. If this is the case, then entanglement is necessary for optimal work extraction.

\section{Extractable work from separable and classical states}

A simple argument shows that separable states, contrary to entangled, do not allow for maximal work extraction. Separable states have the property that the global entropy is greater than all the local entropies \cite{wehrl}. Now, if the system is initially in a separable state $\rho$, then $S(\rho) \geq S(\tau_{\beta})$. This condition, first of all, indicates that the global state cannot be pure \cite{zerotemp}, implying that the bound (\ref{wmax}) cannot be reached by separable states. So, what is the best that classical correlations can do? 

In Appendix~\ref{appa} we show that the locally thermal separable state with the highest extractable work is
\bea
\rho_{\rm sep}=\frac{1}{\mathcal{Z}}\sum_{a=0}^{d-1} e^{-\beta  E_a} \pr{a}^{\otimes n},
\label{StateClas}
\eea
which is simply the state \eqref{e:StateEnt} after being dephased in the (global) energy eigenbasis. Notice that \eqref{StateClas} saturates the inequality $S(\rho) \geq S(\tau_{\beta}(h))$, and in Appendix~\ref{appa2} we show that it is the only separable state with thermal reduced states that saturates it. The extractable work from \eqref{StateClas}, $W_{\rm sep}$, is found, as before, by finding its associated passive state, and then computing the average energy difference, see (\ref{Wmax}). Since $\rho_{\rm sep}$ is already diagonal (with $d$ non-zero eigenvalues), it is only necessary to rearrange these non-zero eigenvalues to the lowest possible energy levels. Let us assume that $n\geq d-1$, (i.e. that we are in the regime of sufficiently many subsystems \cite{fuflfoot}). The $d-1$ largest eigenvalues can then simply be moved into the first excited subspace (with energy $E_1$), giving
\bea \label{wsep}
W_{\rm sep} = nE_{\beta} - E_1(1-\mathcal{Z}^{-1}).
\eea
Note also that $\rho_{\rm sep}$ has no coherences, which means that diagonal and separable states have the same capacity.

\begin{figure}
   \includegraphics[width=8.4cm]{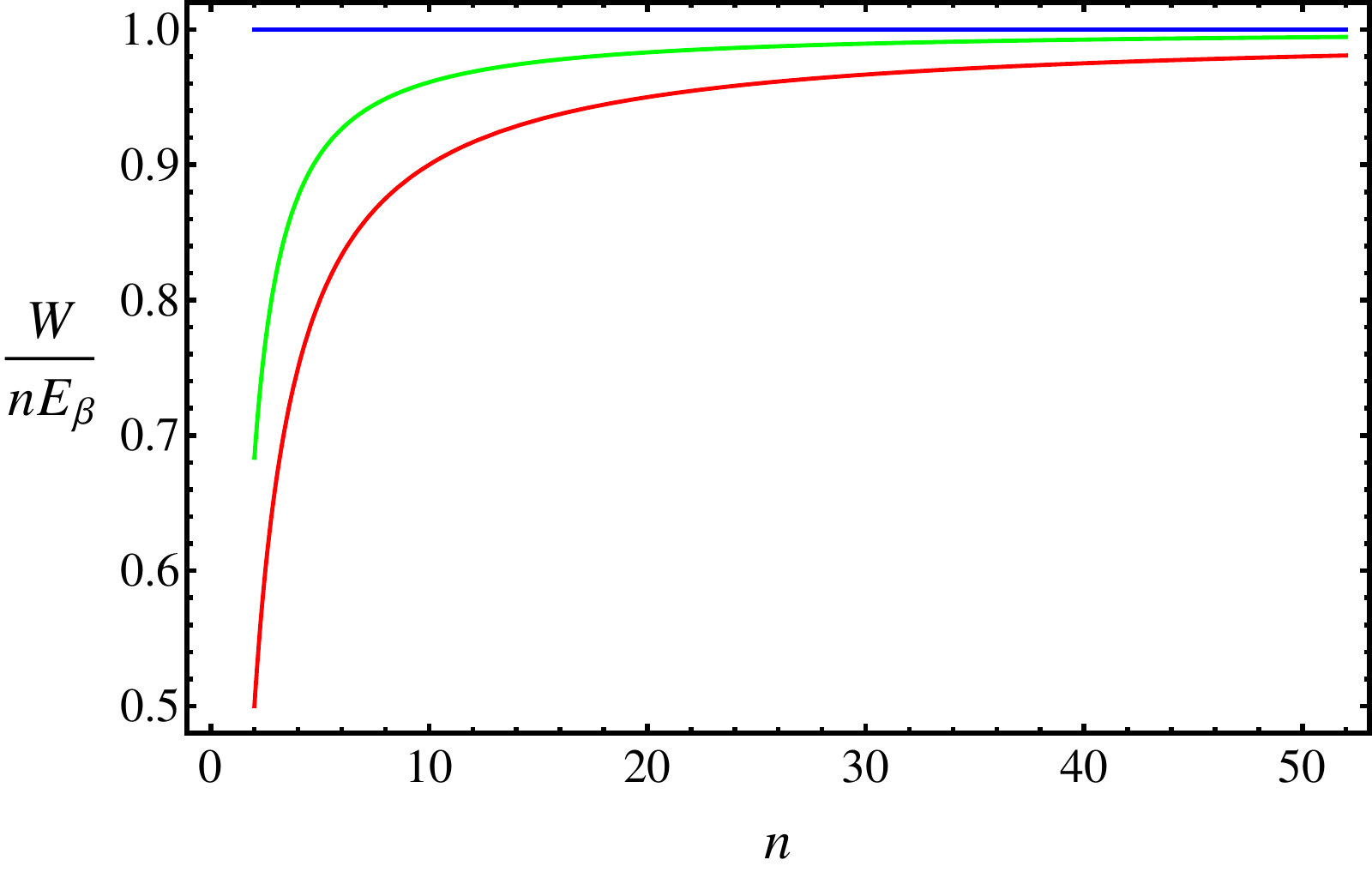}
    \caption{Extractable work from entangled (blue), separable (red), and entangled but having the same entropy as the separable (green) states in units of the initial total energy of the system. Specifically, we take the states \eqref{e:StateEnt}, \eqref{StateClas}, and \eqref{wfixs} for $d = 2$, $\beta E_1=1$. As $n$ increases, classical states become able to store essentially the same amount of work as quantum ones.
}
\label{figI}
\end{figure}

Moreover, as the number of subsystems, $n$, increases, we see that $W_{\rm sep}$ and $W_{\rm max}$ become essentially the same: $W_{\rm sep}/W_{\rm max}=1-\mathcal{O}\left(n^{-1}\right)$ (see Fig.~\ref{figI}). This shows that, in the thermodynamic  limit ($n \rightarrow \infty$), the difference between the extractable work from an entangled state and from a diagonal one vanishes, hence quantum coherences and entanglement play essentially no role here. However, for finite $n$ there will always be a difference. In particular, in the regime of $n$ relatively small, the ability to store work in entanglement offers a significant advantage (see Fig.~\ref{figI}).

At this point let us note that for diagonal initial states (such as (8)), the (average) extractable work as given by the definition \eqref{e:work} coincides with the first moment of work distribution functions introduced in \cite{hanggi,paul}.

\section{Extractable work from states with fixed entropy} \label{sec5}

The previous results can be intuitively understood from entropy considerations. When the correlations in the state are not restricted, it is possible to satisfy the requirement of local thermality with pure entangled states, therefore attaining optimal work extraction. When the state is separable, the global entropy of the state cannot be zero as it is lower bounded by the local entropy and optimal work extraction becomes impossible. Note also that the separable state optimal for work extraction~\eqref{wsep} has global entropy equal to the local one, which means that its global entropy does not scale with the number of subsystems. In other words, its entropy per subsystem tends to zero with the number of subsystems, which intuitively explains why the state tends to be optimal in this limit.

In view of these considerations, it is important to understand how one can store work in correlations when the entropy of the state is fixed. On the one hand, having states whose global entropy scales with the number of subsystems seems more realistic. On the other hand, this allows a more fair comparison between entangled and separable states. In this section we will show that quantum coherences and entanglement enhance the work storage capacity even if the entropy of the global state is fixed. This implies that the entropy gap between separable and entangled states mentioned above is not the only factor making classical states generically worse. However, as in the case of non-restricted entropy, the gain provided by entangled states or energy coherences vanishes in the thermodynamic limit.

Stated otherwise, the question is whether locally thermal quantum states subject to the constraint $S(\rho)=S$ can store more work than (\ref{wsep}) when $S=S(\tau_\beta)$. Now, keeping in mind that local thermality fixes the initial energy to be $nE_\beta$, finding the extractable work, $W_{\rm max}(S)$, amounts to minimizing the final energy, $\tr(H\sigma)$, over all $\sigma= U\rho U^\dagger$ with $U$ being unitary and $\rho$ satisfying the conditions above.

One can readily lower bound $\tr(H\sigma)$ by relaxing all the constraints except $S(\sigma)=S$. Then, as is well known from standard statistical mechanics, the state with the least energy compatible with a given entropy is the thermal state \cite{pusz,lenard}
\begin{equation}
\label{productthermal}
\rho_{\rm th}=\tau_{\beta'}^{\otimes n}
\end{equation}
with $\beta'=\beta'(S)$ being the (unique \cite{yishk}) solution of the entropy constraint $S(\tau_{\beta'})=S/n$. So, $\tr(H\sigma)\geq\tr(H\rho_{\rm th})=n\tr(h\tau_{\beta'})$. This implies a bound on the extractable work
\begin{equation}\label{wfixs} 
W_{\rm max}(S)\leq n E_\beta \left(  1- \frac{1}{E_\beta}  \Tr\left( \tau_{\beta'} h \right) \right).
\end{equation}
In principle, it is not clear if the previous bound is attainable, as the way we found $\rho_{\rm th}$ does not guarantee it to be unitarily achievable from any of the allowed initial states. Nevertheless, as we show below, for any given $S$ and any number $n$ of qubits \cite{ykh} there always exists a locally thermal quantum state that can be transformed to $\rho_{\rm th}$ by a suitable unitary operator, i.e. the bound~\eqref{wfixs} is tight.

Before moving to explicit protocols, let us show a direct consequence of the bound (\ref{wfixs}). As the maximal extractable work from separable states, given in equation \eqref{wsep}, is obtained for $S=S(\tau_{\beta})$, one can easily compare it to $W_{\rm max}(S(\tau_{\beta}))$. The result is illustrated in Fig.~\ref{figI}, showing that $W_{\rm sep}<W_{\rm max}(S(\tau_{\beta}))$. Therefore, even if the entropy is fixed, classical states are generically weaker than entangled quantum states in terms of work storage as the states delivering $W_{\rm max}(S(\tau_{\beta}))$ are necessarily entangled. To understand the reason for this difference, notice that the separability condition restricts the set of locally thermal states (see Appendix~\ref{appa2}), thereby limiting their possible spectra, which, in turn, determine (according to Eq.~\eqref{Wmax}) the extractable work.

Now, let us show an explicit protocol that delivers (\ref{wfixs}). Since to reach the bound in (\ref{wfixs}) the system has to necessarily end up in the state (\ref{productthermal}), we, for clarity, construct the backwards unitary, which takes the final state $\tau_{\beta'}^{\otimes n}$ to an initial state $\rho$ which is locally thermal, at any temperature $\beta \leq \beta'$. In what follows it will be convenient to introduce the local parameter $z = \bra{0}\tau_\beta \ket{0} - \bra{1}\tau_\beta \ket{1}$, i.e. the ``bias'' of the local (qubit) subsystem in state $\tau_\beta$. It is a monotonic function of the temperature: $z = \tanh (\beta E/2)$ (from now on, we concentrate on qubits and, therefore, drop the index of $E_1$). 

We first consider the simplest case of two qubits. Define the unitary transformation $U_\alpha$ to be a rotation by an angle $\alpha$, $\left(\hspace{-1mm}\begin{array}{cc} \cos\alpha & \sin\alpha \\ -\sin\alpha & \cos\alpha \end{array}\hspace{-1mm}\right)$, in $\{\ket{00},\ket{11}\}$ (the subspace spanned by $\ket{00}$ and $\ket{11}$) and an identity on the rest of the space.

If as an initial state we take $\rho = U_\alpha \tau_{\beta'}^{\otimes 2} U_\alpha^\dagger$, then since $U_\alpha$ only generates coherences in the subspace where both qubits are flipped, it is clear that the reduced state of each qubit is diagonal. A straightforward calculation shows that under the action of $U_\alpha$, the state $\tau_{\beta'}$ (with bias $z'$) transforms to an initial state $\rho$ with bias $z = \cos \left(2\alpha\right) z'$. That is, we can achieve any bias $z$ such that $|z| \leq z'$. As such, the local temperature of the initial state, which is simply given by $\beta = \tfrac{2}{E}\tanh^{-1}(\cos \left(2\alpha\right) z')$, can take any temperature $\beta \leq \beta'$ by an appropriate choice of $\alpha$.

The above protocol can be readily generalised to the case of $n$ qubits. Let us denote by $\iv = i_1 \cdots i_n$ an $n$-bit string, with $|\iv| = \sum_{k} i_k$ being the Hamming weight (number of 1s) of the string. The states $\ket{\iv} = \ket{i_1}\cdots\ket{i_n}$ run over all $2^n$ energy eigenstates of $H$. We also introduce $\ivb$|the bit-wise negation of $\iv$ (i.e. $\ket{\ivb} = \sigma_x^{\otimes n}\ket{\iv}$). As we show in Appendix~\ref{appb}, if we now apply an $U_\alpha$ in each of the subspaces $\{\ket{\iv},\ket{\ivb}\}$ with $0\leq |\iv|<n/2$ \cite{footx}, the resulting state is locally thermal, and, exactly as in the case of two qubits, the local bias $z$ is given by $z=\cos \left(2\alpha\right) z'$. Again, any bias $|z|\leq z'$ and therefore any temperature $\beta\leq\beta'$ can be reached.

Notice that the protocol exploits coherence in all two-dimensional subspaces spanned by $\ket{\iv}$ and $\ket{\ivb}$. We expect these optimal states to be entangled in general and, in fact, they are entangled for the scenario depicted in Figure~\ref{figI}. Moreover, we can also show that in the limit of large $n$ the optimal states are necessarily entangled (see Appendix~\ref{appb1}).

Despite this result, in the thermodynamic limit, the bound (\ref{wfixs}) can always be asymptotically reached by (purely classical) diagonal states. To prove this, we distinguish two qualitatively different situations of the global entropy $S$ being macroscopic ($S\propto n$) and sub-macroscopic ($S/n\to0$). In the latter case, as is detailed in Appendix~\ref{appb2}, the proof is conducted by analysing a generalized version of the state (\ref{StateClas}). Whereas the former case of macroscopic entropy can be treated by a simple adaptation of the above protocol. Specifically, as final state one chooses $\rho_{\rm th}=\tau_{\beta'}^{\otimes n}$ with $S(\tau_{\beta'})=\lim_{n\to\infty}S/n$ and, by applying $U_{\frac{\pi}{2}}$, inverts the populations in one subspace $\{\ket{\iv},\ket{\ivb}\}$ with $|\iv|=k\simeq n e^{-\beta'E}/\mathcal{Z'}$. This changes the bias from $z'$ to $z'-\mathcal{O}(1/\sqrt{n})$. So, by performing $\mathcal{O}(\sqrt{n})$ population inversions, one can approximate any $|z|<z'$ and, hence, any temperature $\beta<\beta'$ (see Appendix~\ref{appb3} for details).

By running the above protocol backwards, one immediately notices that the work extraction from correlations is related to the process of their creation from a product of thermal states. In fact, the problem of correlating the latter states as much as possible for a given amount of invested work is considered in \cite{us,david}. There it is shown that the process is optimal when the final state is locally thermal, which is our starting point here. On the other hand, work extraction becomes optimal when the final state is (globally) thermal. That is, the two processes become the reverse of each other \textit{only} when they are both optimal. This situation is in fact common in thermodynamics. For example, a heat engine working at Carnot efficiency can be seen as an optimal refrigerator running backwards \cite{ll5}.

\section{Extension to other scenarios}

Before concluding, we show how our techniques can be applied to other relevant scenarios again in the context of optimal work storage in correlations. In particular, we consider systems where (i) all moments of the energy distribution are equal to those of a global thermal state and (ii) one has access to a thermal bath.

\subsection{Work from energy coherences}

We first consider states whose diagonal (in the energy eigenbasis) is set to be equal to that of a global thermal state, together with the initial condition of local thermality. More formally, this approach is equivalent to imposing that all moments of the energy distribution are those of the global thermal state: $\tr(H^k \rho)=\tr(H^k \tau_\beta^{\otimes n})$, for all $k$. This contrasts with the previous sections where only the first moment (i.e. the average energy) was fixed by local thermality. Moreover, notice that the entropy of the initial state is here unconstrained.

Focusing again first on the case of $n$ qubits, we consider states which are maximally entangled in every degenerate subspace:
\bea
\rho_{\rm deg} = \sum_{k=0}^n C_n^k p^k (1-p)^{n-k} \pr{D_{n,k}}
\label{CorrDeg}
\eea
where $p=e^{-\beta E}/\mathcal{Z}$, and $\ket{D_{n,k}} \propto \sum_{| \iv|=k} \ket{\iv} $ is the normalized Dicke state of $n$ qubits with $k$ excitations. It is straightforward to verify that the above state satisfies equation \eqref{e:locally thermal} and has the required diagonal elements. 

The passive state associated to \eqref{CorrDeg} can be found as follows. Notice that the state \eqref{CorrDeg} is a mixture of $n+1$ orthogonal states. Therefore the optimal unitary amounts to rotating each of these states to the $n+1$ lowest energy levels one of which is the ground state with zero energy and the other $n$ have energy $E$. Therefore the energy of the transformed state is smaller than $E$, which means that it is possible to extract all the energy contained in the initial state up to a correction of $\mathcal{O}(1)$:
\begin{equation}
W_{\rm deg}= nE_\beta - \mathcal{O}(1)E.
\end{equation}
A similar result holds for the general case of $n$ qudits (see Appendix~\ref{appc}).

An interesting question is whether the state $\rho_{\rm deg}$ features entanglement. Intuition suggests that this may be the case, as large coherences are crucial in this scenario. However, using the techniques developed in \cite{Tura}, we have not been able to witness entanglement for $n\leq 50$. Based on this evidence, it seems that in this case entanglement may not provide an advantage for any number of subsystems.

\subsection{Access to a bath}

Finally, we consider an extended scenario in which the system is no longer isolated and can be put in contact with a bath at the same (local) temperature. Here, we ask what is the maximal work that can be extracted via unitaries acting jointly on the system and the bath. Then it is well known that the extractable work is upper bounded by the difference between initial and thermal free energies:
\bea \label{fed}
W\leq F[\rho]-F[\tau_\beta^{\otimes n}],
\eea 
where $F[\rho]=\tr(H\rho)-\beta^{-1}S(\rho)$ and the inequality can be saturated (e.g. via infinitely slow isothermal processes \cite{toof}) \cite{ll5,esposito,paul,aberg}. 

In the present case, the extractable work from any locally thermal state with entropy $S$ is given by
\bea\label{WsepBath_}
W_{\beta,\,{\rm max}}(S)=\beta^{-1}(nS(\tau_\beta)-S),
\eea
where the expression in parentheses is nothing else but a multipartite generalization of the quantum mutual information. This enforces our argument that the origin of the extractable work are the correlations in the state. The bound (\ref{WsepBath_}) is strictly bigger than (\ref{wfixs}), which is natural, as we consider a larger set of operations. On the other hand, the states (\ref{e:StateEnt}) and (\ref{StateClas}) maximize the right hand side of (\ref{WsepBath_}), i.e. the free energy content is maximal, for entangled and separable states respectively, and thus our previous considerations also hold in this framework.

For the case of extracting work from energy coherences, one can readily use (\ref{WsepBath_}) by computing the entropy of \eqref{CorrDeg}. As $\rho_{\rm deg}$ is a mixture of $n+1$ pure states, its entropy cannot exceed (and, as can easily be shown, actually scales as) $\ln (n+1)$. Therefore, $\rho_{\rm deg}$ allows for storing all work in coherences except for a $\mathcal{O}(\ln n)$ correcting term. We note that this optimal state can not be expressed as a tensor product of many coherent states, a situation which was considered previously in the literature \cite{spekkens,paul}.

Notice that, when given access to a bath, the extractable work only depends on a single global property, namely the free energy of the state, which here reduces to the generalized mutual information \eqref{WsepBath_}. Therefore, the strength of the correlations become the only important property, and not whether they are quantum or not. This is in contrast to our previous results in Sec.~\ref{sec5}. In order to reconcile both results, imagine that a bath at temperature $\beta'$ is attached to our system. Then, the bound (\ref{fed}) (with $\beta$ substituted by $\beta'$) will reduce exactly to (\ref{wfixs}). Therefore we see that separable states can saturate (\ref{wfixs}) when a macroscopic object, i.e. a bath, is available. This corroborates our result in Sec.~\ref{sec5}, namely that in the thermodynamic limit \cite{hajat} the difference between quantum and classical correlations vanishes. 

Our results in this section thus complement a previous study \cite{dahlsten} in a similar setting, and also the works \cite{Ueda,Oppenheim,Zurek,AlickiHorodecki,MMVedral,levtoff,Jevtic}, which, although in a completely different context, also deal with the problem of work extraction from thermal environments utilizing correlations. Finally, it is worth mentioning that when the correlations are not present between subsystems but rather between the system and the bath, they become a source of irreversibility \cite{rodi}.

\section{Conclusions}

In this work we investigated and compared the work storing capacities of quantum and classical correlations. To eliminate all sources of work except correlations, we considered systems which are locally thermal. The latter condition is both necessary and sufficient to ensure that the system becomes passive once the correlations are removed. This gives a new perspective on the problem of passivity, in particular for the case of composite systems. 

We first show that correlations are powerful enough to allow for the extractable work to be equal to all the energy present in the system (see Sec.~\ref{sec3}). For that to happen, the state of the system must not only be entangled but also pure, which is impossible for locally thermal separable states due to an entropy constraint. Entanglement is also useful when the state of the system is mixed, as in this case we show that separable states can not generically store the maximal work compatible with the entropy of the system and local thermality.

Furthermore, we prove that in all cases the quantum advantage, significant for small ensembles, becomes irrelevant in the thermodynamic limit, thereby corroborating that macroscopic thermodynamics is insensitive to the microscopic mechanics underlying it. This ``classical'' view is complemented by a previous result by some of us \cite{karen} stating that maximal work can be extracted from diagonal states without generating entanglement during the whole process.

The considered scenario, a set of correlated yet locally thermal states, is ideal to identify the role of quantum effects in thermodynamics and naturally allows for extensions. In this respect, first we have studied the role of coherences by further restricting the diagonal of the state in the energy eigenbasis to be identical to a thermal state. Interestingly, in this case it turns out that, in the thermodynamic limit, essentially all the energy can be stored in the off-diagonal terms. Secondly,  we have discussed the situation when the system is allowed to interact with a thermal bath at the local temperature of the reduced states. Then, work is directly related to the strength of the correlations as measured by \eqref{WsepBath_}.

An interesting open question is to investigate the scenario in which not only local marginals are thermal, but so are also $k$-body reduced states (in particular the case of nearest neighbours). This may give an insight into the role of different types of multipartite entanglement in the context of work extraction. Another interesting question is to derive bounds in the other direction, i.e., correlated states with minimal work content \cite{usIII}. A promising line of further research is to study the process of converting correlations into work beyond average quantities, for example from the point of view of fluctuation theorems \cite{hanggi}, or deterministic work extraction  \cite{horodecki,aberg,mueller}.


\acknowledgements

We gratefully acknowledge discussions with Jordi Tura. This work is supported by the IP project SIQS, the Spanish project FOQUS, and the Generalitat de Catalunya (SGR 875). M. P. L. acknowledges funding from the Severo Ochoa program and the Spanish grant FPU13.05988, M.H. from Marie Curie Grant No. 302021 “Quacocos”, by the European Commission (STREP ``RAQUEL''), the Spanish MINECO, projects FIS2008-01236 and FIS2013-40627-P, with the support of FEDER funds, by the Generalitat de Catalunya CIRIT, project 2014-SGR-966 and the Juan de la Cierva fellowship (JCI 2012-14155)., N. B. from the Swiss National Science foundation (Grant No. PP00P2.138917) and  SEFRI (COST action MP1006), P.S. from the Marie Curie COFUND action through the ICFOnest program, and the ERC AdG NLST, A.A. is supported by the ERC CoG QITBOX, and all authours thank the EU COST Action MP1209 ``Thermodynamics in the quantum regime". At initial stages, parts of this work were carried out at the Quantum Information 2013 conference at Benasque (Spain) and the Information theoretic approaches to thermodynamics conference at The Institute for Mathematical Sciences, NUS (Singapore).

\appendix

\section{Maximal work from separable states} \label{appa}

In this appendix we find the maximal work that can be stored in separable states subject to being locally thermal.

\subsection{The set} \label{appa1}

In other words, we want to find the maximum of the ergotropy (\ref{Wmax}),
\bea \label{Wmax2}
W_{\rm max}(\rho)=\tr(H\rho)-\tr(H\rho^{\rm passive}),
\eea
over all those $\rho$s that belong both to the set of separable states (we denote it as $\mathcal{SEP}$) and to 
\bea
\mathcal{LTH}=\left\{\rho\, : \, \tr_{\overline{i}}\rho=\tau_\beta,\; i=1,...,n\right\}.
\eea
Now observe that, along with $\mathcal{SEP}$ \cite{zhuchok}, $\mathcal{LTH}$ is a convex set. Indeed, if $\rho_1$ and $\rho_2$ are arbitrary two states belonging to $\mathcal{LTH}$, then for $\forall  t\in[0,1]$
\bea
\tr_{\overline{i}}(t\rho_1+(1-t)\rho_2)=t\tau_\beta+(1-t)\tau_\beta=\tau_\beta
\eea
for all $i$, immediately implying that $t\rho_1+(1-t)\rho_2\in\mathcal{LTH}$ for all $t\in[0,1]$ which, by definition, means $\mathcal{LTH}$ is a convex set. Moreover, since the conditions defining $\mathcal{LTH}$ are linear, it is also a closed set. 

We will need also the following set:
\bea
\mathcal{ENT}(S)=\left\{\rho\, : \, S(\rho)\geq S\right\}.
\eea
Due to convexity of the von Neumann entropy, $\mathcal{ENT}$ is also convex and the not strict inequality in the definition ensures that it is also closed.

Another observation is that since the entropy of the separable states is greater than all the local entropies, we have that if $\rho\in\mathcal{SEP}\cap\mathcal{LTH}$ then $S(\rho)\geq S(\tau_\beta)$. Otherwise:
\bea \label{puc1}
\mathcal{SEP}\cap\mathcal{LTH}\subset \mathcal{ENT}(S(\tau_\beta)).
\eea
Moreover, the intersection of the boundaries of all three sets in (\ref{puc1}) is nonempty and consists of only one element which we find in the next subsection.

\subsection{Maximally pure separable state} \label{appa2}

Here we determine the separable state $\rho_{\rm sep}$ of $N$ systems (all having the same $d$ level Hamiltonian $h=\sum_{a=0}^{d-1}E_a |a\rangle\langle a|$) such that it has the minimal entropy compatible with marginals all being $\tau_\beta$ (\ref{tb}):
\bea \label{tau}
\tau_\beta=\frac{1}{Z}e^{-\beta h}=\frac{\sum\limits_{a=0}^{d-1} e^{-\beta E_a} |a\rangle\langle a|}{\sum\limits_{b=0}^{d-1} e^{-\beta E_b}}\equiv\sum_{a=0}^{d-1} p_a |a\rangle\langle a|.~~~
\eea
Considering, e.g., the first system ($S^1$) versus the rest ($R=S^2\otimes \cdots\otimes S^N$) and keeping in mind that the partial states of $S$s are all $\tau_\beta$, we have
\bea \label{puc}
S(\rho_{\rm sep})-S(\tau_\beta)=-S\left(\rho_{\rm sep}\;\big|\big|\;\tau_\beta\otimes\frac{I_R}{d_R}\right)+\ln d_R.~~~~~
\eea
Since $\rho_{\rm sep}$ is separable, it can be written in the following form:
\bea
\rho_{\rm sep}\hspace{-0.25mm}=\hspace{-0.25mm}\sum_x \lambda_x \rho^{S^1}_x\hspace{-0.5mm}\otimes\rho^R_x=\hspace{-0.25mm}\sum_x \lambda_x \rho^{S^1}_x\hspace{-0.5mm}\otimes\rho^{S^2}_x\hspace{-0.5mm}\otimes\cdots\otimes\rho^{S^N}_x~~~~~~
\eea
for some discrete index $x$, nonnegative $\lambda_x$s summing up to $1$, and some normalised states $\rho^{S^i}_x$ over $S^i$. Given the condition that the state of $S^1$, $\sum_x \lambda_x \rho^{S^1}_x$, is equal to $\tau_\beta$ and the joint convexity of the relative entropy \cite{JR}, we have
\begin{align}
S(\rho_{\rm sep})-S(\tau_\beta) = 
\nonumber\\
\ln d_R - S\left(\sum_x \lambda_x \rho^{S^1}_x \otimes\rho^R_x \;\big|\big|\; \sum_x \lambda_x \rho^{S^1}_x \otimes \frac{I_R}{d_R} \right)\geq \nonumber
\\ 
 \ln d_R - \sum_x \lambda_x S\left(\rho^{S^1}_x \otimes\rho^R_x \;\big|\big|\; \rho^{S^1}_x \otimes \frac{I_R}{d_R} \right)=
\nonumber\\
 \sum_x \lambda_x S(\rho^R_x)\geq 0.
 \label{ineqs}
\end{align}
So, the minimal possible value for $S(\rho_{\rm sep})$ is $S(\tau_\beta)$; and to find the purest $\rho_{\rm sep}$ we have to saturate both inequalities in the chain (\ref{ineqs}). The second inequality is resolved trivially, giving that $\rho^R_x=\rho^{S^2}_x\otimes\cdots\otimes\rho^{S^N}_x$ for all values of $x$ are pure. We denote these states as $|R_x\rangle=|S^2_x\rangle\otimes\cdots\otimes|S^N_x\rangle$. Doing the same with respect to, e.g., $S^2$, we will get that all $\rho_x^{S^1}$ are also pure (and, as above, are denoted as $|S^1_x\rangle$).

The equality conditions for the first inequality of (\ref{ineqs}) are less trivial \cite{JR}. If we only consider the nonzero $\lambda_x$s and denote their number by $L$, {\it Theorem 8} of \cite{JR} will give us
\begin{align} \label{condi1}
\left(\lambda_x \rho^{S^1}_x \otimes\rho^R_x\right)^{it}\left(\lambda_x \rho^{S^1}_x \otimes\frac{I_R}{d_R}\right)^{-it}=\rho_{\rm sep}^{it}\left(\tau_\beta \otimes\frac{I_R}{d_R}\right)^{-it}
\nonumber\\
\text{for}\quad \forall t>0 \quad \text{and}\quad x=0,...,L-1;
\end{align}
where the equality holds in the support of $\rho_x^{S^1}\otimes\rho_x^R=|S^1_x\cdots S^N_x\rangle\langle S^1_x\cdots S^N_x|=|S_x\rangle\langle S_x|=P_x$ (in this notation $\rho_{\rm sep}=\sum_x\lambda_x P_x$). The latter is the projector onto that subspace. Bearing in mind that we consider only nonzero $\lambda_x$s and doing the same procedure for all other $N-1$ systems, we get from (\ref{condi1}):
\bea \label{preultim}
P_x\rho_{\rm sep} P_x=P_x (\tau_\beta\otimes I_{S^2}\otimes\cdots\otimes I_{S^N})P_x = \cdots =~~~~~~~~~~ 
\nonumber\\
=P_x (I_{S^1}\otimes I_{S^2}\otimes\cdots\otimes \tau_\beta)P_x.~~~~~~~~
\eea
We will now concentrate on the first equality and, for simplicity, drop the index enumerating the subsystems. With that, and taking into account that $P_x\rho_{\rm sep} P_x=\lambda_x P_x$ and $P_x (\tau_\beta\otimes I_{S^2}\otimes\cdots\otimes I_{S^N})P_x=\langle S_x|\tau_\beta|S_x\rangle P_x$, we have
\bea \label{lalala}
\lambda_x=\langle S_x|\tau_\beta|S_x\rangle.
\eea
Now we take $\{|a\rangle\}_{a=0}^{d-1}$, the eigenbasis of $\tau_\beta$ in the Hilbert space of the subsystem (\ref{tau}), and construct the matrix $m_{xa}=\left| \langle S_x| a\rangle \right|^2\geq0$. With this we rewrite (\ref{lalala}) as
\bea \label{1}
\sum_{a=0}^{d-1} m_{xa}p_a=\lambda_x.
\eea
Also, from the normalization we have
\bea \label{2}
\sum_a m_{xa}=1 \quad \text{for}\quad \forall x.
\eea
Finally, the condition that all partial states are $\tau_\beta$: $\sum_x \lambda_x |S_x\rangle\langle S_x|=\tau_\beta$, leads us to
\bea \label{3}
\sum_{x=0}^{L-1} \lambda_x m_{xa}=p_a.
\eea
First, let us show that $L>d$ cannot be true. Indeed, substitute (\ref{1}) into (\ref{3}), $\sum_{xb}m_{xa}m_{xb}p_b=p_a$, multiply the LHS by $m_{xa}$ and sum over $a$ and use $\sum_x\lambda_x=1=\sum_{xa}m_{xa}p_a$:
\bea \label{hajat}
\sum_x \left(\sum_a m_{xa}^2\right)\left( \sum_b m_{xb}p_b \right)=1.
\eea
Given that it must hold that $\sum_{xa}m_{xa}p_a=1$ we see that (\ref{hajat}) can be true only if
\bea \label{pupul}
\sum_{a}m_{xa}^2=1\quad \text{for}\quad \forall x.
\eea
But we have (\ref{2}) and that $0\leq m_{xa}\leq 1$ so (\ref{pupul}) can be true only if each row consists of zeroes and only one $1$. Since none of $p_a$ is zero, (\ref{3}) implies that there must be at least one $1$ on each column of $m$. Let arrange the $x$ so that the first $d$ rows of $m$ look like an identity matrix. Then we get
\bea \label{4}
\lambda_x=p_x \quad \text{for}\quad x=0,...,d-1.
\eea
Since $\sum_x\lambda_x=1$ we have that $\lambda_x=0$ for all $x\geq d+1$. Which is impossible because of (\ref{1}) and the fact that there must be at least one $1$ on each row.

With the same argument, also $d>L$ is not possible. So, $d=L$ and (\ref{4}) holds. Also, since now $m=I$, $|S_x\rangle=|x\rangle$, rendering
\bea \label{puc55}
\rho_{\rm sep}=\sum_{a=0}^{d-1} p_a \pr{a\cdots a}.
\eea
Moreover, since $\rho_{\rm sep}$ is unique,
\bea \label{puc6}
\partial\mathcal{ENT}(S(\tau_\beta))\cap\mathcal{SEP}\cap\mathcal{LTH}=\{\rho_{\rm sep}\},
\eea
where $\partial$ denotes the boundary of the set.

\subsection{Convexity of ergotropy} \label{appa3}

In this section we take another step towards finding the maximum of the ergotropy $W(\rho)$ (\ref{Wmax},\,\ref{Wmax2}) over $\mathcal{SEP}\cap\mathcal{LTH}$. To that end we prove a general result which does not depend on the particular structure of the system we discuss in this article: {\it on the set of states with equal energy, ergotropy is a convex function}.

So, say we are given the Hamiltonian $H=\sum_\alpha\mathcal{E}_\alpha\pr{\alpha}$ with $\mathcal{E}_{\alpha+1}\geq\mathcal{E}_\alpha$, $\alpha=0,1,...$. Now, for any $\rho_1$ and $\rho_2$ st $\tr(H\rho_1)=\tr(H\rho_2)$ and $\forall t\in[0,1]$,
\bea \label{puc1}
W_{max}(t\rho_1+(1-t)\rho_2)\leq tW_{max}(\rho_1)+(1-t)W_{max}(\rho_2).~~~~
\eea
To prove this, observe that (\ref{puc1}) is equivalent to
\bea\label{puc3}
\tr(H [t\rho_1+(1-t)\rho_2]^{\rm passive})\geq \\
 t\tr(H\rho_1^{\rm passive})+(1-t)\tr(H\rho_2^{\rm passive})= \\ \label{puc4} 
 \tr(H[t\rho_1^{\rm passive}+(1-t)\rho_2^{\rm passive}]).
\eea
On the other hand, as is shown in \cite{armen}, for two diagonal states $\rho$ and $\sigma$,
\bea \label{puc5}
\rho\prec\sigma\; \Rightarrow\; \tr(H\rho)\geq\tr(H\sigma),
\eea
where $\rho\prec\sigma$ is read as $\rho$ {\it is majorized by} $\sigma$ and means that
\bea
\sum_{\alpha=0}^A\rho_{\alpha\alpha}\leq\sum_{\alpha=0}^A\sigma_{\alpha\alpha},\;\text{for all}\;A=0,1,....
\eea
Now, as a direct consequence of the Theorem G.1. of chapter 9 of \cite{molkin}, we have
\bea
[t\rho_1+(1-t)\rho_2]^{\rm passive}\prec t\rho_1^{\rm passive}+(1-t)\rho_2^{\rm passive},~~~~~~~
\eea
which, in view of (\ref{puc5}), leads to (\ref{puc3},\,\ref{puc4}), which prove (\ref{puc1})|the main result of this subsection.

\subsection{Maximization of work over $\mathcal{SEP}\cap\mathcal{LTH}$} \label{appa4}

We are now ready to prove the main claim of this section, namely: 
\bea \label{puc7}
\max_{\rho\in\mathcal{SEP}\cap\mathcal{LTH}}W_{max}(\rho)=W_{max}(\rho_{\rm sep}),
\eea
where $\rho_{\rm sep}$ is from (\ref{puc55}).

Consider the set
\bea
\Sigma(S)=\mathcal{ENT}(S)\cap\mathcal{SEP}\cap\mathcal{LTH}.
\eea
As a union of closed convex sets, $\Sigma$ is a closed convex set. Eq. (\ref{puc1}) implies that $\Sigma(S(\tau_\beta))=\mathcal{SEP}\cap\mathcal{LTH}$.   Also, obviously, when $S(\rho)>nS(\tau_\beta)$, $\rho$ cannot be in $\mathcal{LTH}$ and therefore $\Sigma(S)=\emptyset$ for all $S>nS(\tau_\beta)$, and $\Sigma(nS(\tau_\beta))=\{\tau_\beta^{\otimes n}\}$.

A convex function has its maximum over a closed convex set on the boundary (more precisely on one of the extremal points) of that set \cite{rock}. Now, since all $\rho$s in $\mathcal{SEP}\cap\mathcal{LTH}$ are by definition locally thermal, they all have the same energy $\tr(H\rho)=nE_\beta$, which, according to the previous subsection, ensures that $W(\rho)$ is a convex function on the whole set $\mathcal{LTH}$. Moreover, it has its maximum, $\mathcal{W}(S)$, over $\Sigma(S)$ on $\partial\Sigma(S)$. Also, since this maximum changes with $S$, the point delivering it lies on the boundary of $\mathcal{ENT}(S)$. On the other hand, since $\Sigma(S_1)\subset\Sigma(S_2)$ when $S_1>S_2$, then $\mathcal{W}(S_1)<\mathcal{W}(S_2)$. Finally, as $\mathcal{W}(S)$ is a monotonically decreasing function of the global entropy, it has its maximal value at $S=S(\tau_\beta)$|the minimal possible entropy. Furthermore, because $\Sigma(S(\tau_\beta))=\mathcal{SEP}\cap\mathcal{LTH}$, 
\bea
\mathcal{W}(S(\tau_\beta))=\max_{\rho\in\mathcal{SEP}\cap\mathcal{LTH}}W_{max}(\rho),
\eea
and this maximum is attained on the boundary of $\mathcal{ENT}(S(\tau_\beta))$. Since the latter intersects $\mathcal{SEP}\cap\mathcal{LTH}$ in only one point, $\rho_{\rm sep}$ (see (\ref{puc6})), means that the latter is the point where $W(\rho)$ attains its maximal value, which proves (\ref{puc7}).

\section{Protocol for maximal work extraction given an entropy constraint}\label{appb}

In this appendix we will show that the unitary $U_{\boldsymbol{\alpha}}$, with ${\boldsymbol{\alpha}} = \alpha \cdots \alpha$, given by
\begin{align}
U_{\boldsymbol{\alpha}} \ket{\iv} &= \cos\alpha \ket{\iv} + \sin\alpha\ket{\ivb}, & \bra{\iv}H_0\ket{\iv} < \tfrac{n}{2}\nonumber \\
U_{\boldsymbol{\alpha}} \ket{\ivb} &= -\sin\alpha \ket{\iv} + \cos\alpha\ket{\ivb}, & \bra{\iv}H_0\ket{\iv} < \tfrac{n}{2} \\
U_{\boldsymbol{\alpha}} \ket{\iv} &=  \ket{\iv}, & \bra{\iv}H_0\ket{\iv} = \tfrac{n}{2} \nonumber
\end{align}
produces a state $\rho = U_\alpha \tau_{\beta'}(H_S)^{\otimes n} U_\alpha^\dagger$ that is locally thermal with local bias $z$ and temperature $\beta$ given by
\begin{align}
z &= \cos \left(2\alpha\right) z'\\
\beta &= \nonumber \tfrac{2}{E}\tanh^{-1}(\cos \left(2\alpha\right) z')
\end{align}
where $z' = \bra{0}\tau_{\beta'} \ket{0} - \bra{1}\tau_{\beta'} \ket{1} = \Tr (\sigma_z \tau_{\beta'})$ is the bias of $\tau_{\beta'}$ (where, for the sake of brevity, we now write $\tau_{\beta'}$ in place of $\tau_{\beta'}(H_S)$ since no confusion should arise).
To see that this is the case, we note first that $\rho$ is symmetric under permutations, since both the initial state $\tau_{\beta'}(H_S)^{\otimes n}$ and $U_\alpha$ are symmetric. Therefore it suffices to calculate $z_1 = \bra{0}\rho_1 \ket{0} - \bra{1}\rho_1 \ket{1}$. We  note first that this can be re-written as follows
\begin{align}
    z_1 &= \Tr \left(\sigma_z \rho_1 \right) = \Tr\left(\sigma_z \otimes \openone_{n-1}\rho\right) \nonumber \\
    &= \sum_{i_1\cdots i_n} \bra{\mathrm{i}}(-1)^{i_1} \rho \ket{\mathrm{i}} \nonumber
\end{align}
Now, it is straightforward to see that
\begin{align}
\bra{\mathrm{i}}\rho \ket{\mathrm{i}} &= \bra{\mathrm{i}}U_\alpha\tau_{\beta'}^{\otimes n} U_\alpha^\dagger\ket{\mathrm{i}} \nonumber \\
&= \cos^2\alpha \bra{\mathrm{i}}\tau_{\beta'}^{\otimes n} \ket{\mathrm{i}} + \sin^2\alpha \bra{\bar{\mathrm{i}}}\tau_{\beta'}^{\otimes n} \ket{\bar{\mathrm{i}}}
\end{align}
holds for all $\ket{\mathrm{i}}$, and futhermore that $\bra{i} \tau_{\beta'} \ket{i}=\tfrac{1}{2}(1 + (-1)^{i}z')$, which follows from the definition of $z'$ as the bias. Put together, this allows one to re-express $z_1$ as
\begin{multline}
z_1 = \sum_{i_1\cdots i_n} (-1)^{i_1} \left(\frac{\cos^2 \alpha}{2^n}\prod_k (1+(-1)^{i_k}z') \right. \\
\left. + \frac{\sin^2 \alpha}{2^n}\prod_k (1+(-1)^{i_k}(-z'))\right)
\end{multline}
which, upon interchanging the order of the product and sum becomes
\begin{multline}
z_1 = \frac{\cos^2 \alpha}{2^n}\prod_{i_1\cdots i_n} \sum_{i_k} (-1)^{i_1}(1+(-1)^{i_k}z') \\
+ \frac{\sin^2 \alpha}{2^n}\prod_{i_1\cdots i_n} \sum_{i_k} (-1)^{i_1}(1+(-1)^{i_k}(-z'))
\end{multline}
For $k \neq 1$, $\sum_{i_k} (-1)^{i_1}(1+(-1)^{i_k}z') = 2$, whilst for $k = 1$, $\sum_{i_k} (-1)^{i_1}(1+(-1)^{i_k}z') = 2z'$, from which we finally obtain
 \begin{align}
z_1 &= \cos^2 (\alpha) z'+ \sin^2 (\alpha)(-z') \nonumber \\
&= \cos(2\alpha)z'
\end{align}

\subsection{Presence of entanglement in the state} \label{appb1}
Consider the state $\rho=U_{\alpha} \tau_{\beta'}^{\otimes n} U_{\alpha}^{\dagger}$\comments{, where each $\tau_{\beta'}$ is a qubit with population at the excited state $p=e^{-\beta' \epsilon}/\mathcal{Z}$}. As it has an X-like shape, applying the criterion of positivity under partial transposition (PPT)  \cite{marcusX,marcusY} with respect to a bipartition $A|\bar{A}$ to $\rho$ will yield an independent positivity condition for each pair of coherences $\langle  {\bf i} |\rho | \bar{{\bf i}} \rangle$, $\langle \bar{{\bf i}} |\rho | {\bf i} \rangle$, given by
\begin{eqnarray}
 | \bra{{\bf i}} \rho \ket{{\bf \bar{i}}} | - \sqrt{\bra{{\bf i}} \bra{{\bf \bar{i}}} \Pi_{A|\bar{A}} \rho^{\otimes 2} \Pi_{A|\bar{A}} \ket{{\bf i}} \ket{{\bf \bar{i}}}}  \geq 0
\end{eqnarray}
where $\Pi_{A|\bar{A}}$ is the permutation operator acting on the two-copy Hilbert space exchanging partition $A$ between the two copies.
 Focusing on $|{\bf i} \rangle = | 0...0 \rangle$, $|\bar{{\bf i}} \rangle =|1...1 \rangle$ and on the bipartition $(n/2|n/2)$, the condition for non-separability reads:
\begin{equation}\label{b8}
\sin (2\alpha) (1-e^{-\beta' \epsilon n})-2 e^{-\beta' \epsilon n/2} \geq 0.
\end{equation}
For sufficiently large $n$, entanglement will be present in the state for any $\alpha$. Indeed, when $S(\rho)\propto n$, $\beta'$ is a constant, and so is $\alpha$. So, for $n$ large enough, the LHS of (\ref{b8}) will be $\approx \sin(2\alpha)$ which is $\geq 0$. In all other cases, i.e. when $S(\rho)\not\propto n$, which means $S(\tau_{\beta'})=\frac{S(\rho)}{n}\to0$ (with $n\to\infty$), $e^{-\beta'\epsilon}$ decreases, so $z'=\frac{1-e^{-\beta'\epsilon}}{1+e^{-\beta'\epsilon}}$ increases, so $\cos(2\alpha)=\frac{z}{z'}$ decreases, so $\sin(2\alpha)$ increases. All in all, the LHS of (\ref{b8}) increases with $n$, becoming positive starting from some value of $n$.

\subsection{Maximal work extraction from states with {\it sub}macroscopic entropy}\label{appb2}

Here we show that when the entropy of the global state, $S(\rho)$, is sub-macroscopic, i.e. 
\bea
x_n=\frac{S(\rho)}{n}\to 0\quad\text{when}\quad n\to\infty,
\eea
The maximal works extractable from locally thermal separable states, $W_{\rm sep}(S(\rho))$, and from general entangled locally thermal states, $W_{\rm max}(S(\rho))$ (\ref{wfixs}), asymptotically coincide:
\bea\label{mrappb2}
\lim\limits_{n\to\infty}\frac{W_{\rm sep}(S(\rho))}{W_{\rm max}(S(\rho))}=1.
\eea
First we observe that, trivially,
\bea
\frac{W_{\rm sep}(S(\rho))}{W_{\rm max}(S(\rho))}\leq1
\eea
We then start by asymptotically expanding $W_{\rm max}(S(\rho))$. For that we will need the asymptotics of $E_{\beta'}$ when $S(\tau_{\beta'})=x_n$. Denote $p'=e^{-\beta'E}/\mathcal{Z}'$. Then $E_{\beta'}=p'E$. Now, since $x_n\to0$, $p'$ also has to $\to0$. Therefore 
\bea \nonumber
x_n=-p'\ln p'-(1-p')\ln(1-p')=p'\ln\frac{1}{p'}+\mathcal{O}(p').
\eea
Hence,
\bea
p'=\frac{x_n}{\ln\frac{1}{x_n}}\left[ 1+\mathcal{O}\left( \frac{\ln\ln\frac{1}{x_n}}{\ln\frac{1}{x_n}} \right) \right].
\eea
And since the final energy is simply $nE_{\beta'}=np'E$, we have
\bea\label{wmaxappb2}
W_{\rm max} = nE_\beta-\frac{S}{\ln n-\ln S}(1+o(1))\sim nE_\beta.
\eea

Let us now consider the following three parameter family of diagonal states:
\bea \label{pnto0}
\Omega(\epsilon,\delta,\gamma)=\epsilon\pr{0}^{\otimes n}+\delta\pr{1}^{\otimes n}+\frac{\gamma}{C_n^D}\sum_{|\iv|=D}\pr{\iv},~~~~~~~~
\eea
where $D$ is the smallest number satisfying $\ln C_n^D\geq S$, and $\epsilon$, $\delta$, and $\gamma$ are nonnegative and, from normalization condition:
\bea \label{pnto1}
\epsilon+\delta+\gamma=1.
\eea
Furthermore, the local thermality requires
\bea \label{pnto2}
\delta+\gamma\frac{D}{n}=\frac{e^{-\beta E}}{\mathcal{Z}}\equiv p.
\eea
And finally, the entropy must be $S$:
\bea \label{pnto3}
-\epsilon\ln\epsilon-\delta\ln\delta-\gamma\ln\gamma+\gamma\ln C_n^D=S.
\eea
Resolving (\ref{pnto1}) and (\ref{pnto2}) we reformulate (\ref{pnto3}) as
\bea \label{pnto}
f(\gamma)=S
\eea
where 
\bea\nonumber
f(\gamma)=&-&\left(1-p-\gamma \frac{n-D}{n}\right)\ln\left(1-p-\gamma \frac{n-D}{n}\right) \\ \nonumber &-& \left(p-\gamma \frac{D}{n}\right)\ln\left(p-\gamma \frac{D}{n}\right)-\gamma\ln\gamma+\gamma\ln C_n^D.~~~~
\eea
Now, $f(\gamma)$ is a continuous function on $[0,1]$, and $f(0)=-p\ln p-(1-p)\ln(1-p)=S(\tau_\beta)\leq S$ and $f(1)=\ln C_n^D$ which, by the very definition of $D$, exceeds $S$. Note also, that local thermality places an upper bound on $S$: $S\leq $ So, $S\in[f(0),f(1)]$ and, due to the continuity of $f(\gamma)$, $\exists \gamma$ such that (\ref{pnto}) is satisfied. We denote that value of $\gamma$ via $\gamma_0$, and the state $\Omega$ it (uniquely) determines|via $\Omega_0$.

Finally, we note that since the rank of $\Omega_0$ is at most $2+C_n^D$, the passive state associated to it will occupy the first $2+C_n^D < C^{D+1}_n$ energy levels. Therefore the energy of $\Omega_0$ is $< (D+1)E$. Hence
\bea
W_{\rm sep}> nE_\beta-(D+1)E.
\eea
On the other hand, $D<S$. Indeed, due to the general inequality $C_n^k\geq \left(\frac{n}{k}\right)^k$, we have $\ln C_n^S\geq S\ln\frac{n}{S}$. Since $S/n\to 0$, for sufficiently big $n$ we will have $n/S>e$, which yields to $\ln C^S_n>S$ implying that $D<S$. Thereby, we end up with
\bea
W_{\rm sep}>nE_{\beta}-SE,
\eea
which, taking into account (\ref{wmaxappb2}), leaves us with
\bea\label{horom}
\frac{1-\frac{S}{n}\frac{E}{E_\beta}}{1-\frac{S}{n}\frac{1+o(1)}{\ln n-\ln S}}<\frac{W_{\rm sep}(S(\rho))}{W_{\rm max}(S(\rho))}\leq1.
\eea
In view of $S/n\to 0$, (\ref{horom}) finalizes the proof of (\ref{mrappb2}).

\subsection{Maximal work extraction from states with macroscopic entropy}\label{appb3}

In what follows we will show that in the asymptotic limit it is possible to approximately achieve maximal work extraction given an entropy constraint from a state which is classical. To do so we shall apply the unitary $U_{\boldsymbol{\alpha}}$ with $\boldsymbol{\alpha}$ chosen appropriately. Consider that $\alpha_k$ is non zero (and equal to $\pi/2$) only for $k = np' - \mu\equiv \ell$, i.e., between the subspaces with $|\iv| = np' - \mu$ and $|\ivb| = n(1-p') + \mu$, where $p' = \bra{1}\tau_{\beta'}\ket{1} = \tfrac{1}{2}(1-z')$ is the excited state probability in $\tau_{\beta'}$. That is, we consider the unitary $V$
\bea
V \ket{\iv} &=  \ket{\ivb}, \; V\ket{\ivb}=-\ket{\iv} &  \quad\quad\text{if} \quad\quad  |\iv| = np'-\mu\nonumber \\
V \ket{\iv} &=  \ket{\iv} \phantom{, \; V\ket{\ivb}=-\ket{\iv}} & \quad\quad\text{if} \quad\quad |\iv|\neq np'-\mu. \nonumber
\eea
Obviously, after applying $V$ the state is still diagonal and symmetric. This means that the transformed state is again locally thermal, but now with the new bias $z''=1-2p''$. To find it, we observe that the energy of the global state is given by $nEp''$. On the other hand, $V$ swapped the population of the level $\ell E$, $C_n^{\ell}(p')^{\ell}(1-p')^{n-\ell}$, with $C_n^{n-\ell}(p')^{n-\ell}(1-p')^{\ell}$, the population of $(n-\ell)E$. As a result, the initial energy $np'E$ increased by $C_n^{\ell}\left( (p')^{\ell}(1-p')^{n-\ell} -(p')^{n-\ell}(1-p')^{\ell} \right)(n-2\ell)E$. This implies, that
\bea \nonumber
p'' &=& p' \\ \nonumber &+& C_n^{\ell}\left( (p')^{\ell}(1-p')^{n-\ell} -(p')^{n-\ell}(1-p')^{\ell} \right)(1-2\ell/n),
\eea
or, equivalently,
\bea \nonumber
z''\hspace{-0.5mm} &=& \hspace{-0.5mm} z' \\ \nonumber &-& \hspace{-0.5mm} 2C_n^{\ell}(z'+2\mu/n)\left( (p')^{\ell}(1\hspace{-0.5mm}-\hspace{-0.5mm}p')^{n-\ell}\hspace{-0.5mm} -(p')^{n-\ell}(1\hspace{-0.5mm}-\hspace{-0.5mm}p')^{\ell} \right). \\ \label{procakant}
\eea

Now, let us focus on $\mu\leq\mathcal{O}(\sqrt{n})$ (we will see that this set is enough for our purposes). We then have the asymptotic expansion
\bea\label{banadzev}
(\hspace{-0.25mm}p'\hspace{-0.25mm})^{np'\hspace{-0.25mm}-\hspace{-0.25mm}\mu} (\hspace{-0.25mm}1\hspace{-0.5mm}-\hspace{-0.5mm}p'\hspace{-0.25mm})^{n(\hspace{-0.2mm}1\hspace{-0.25mm}-\hspace{-0.25mm}p'\hspace{-0.2mm})+\mu} C_{n}^{np'-\mu} \hspace{-0.75mm}=\hspace{-0.75mm} \frac{e^{-\frac{\mu^2}{2p'(\hspace{-0.25mm}1\hspace{-0.25mm}-\hspace{-0.25mm}p'\hspace{-0.25mm})n}+\mathcal{O}\left(\hspace{-0.5mm}\frac{\mu}{n}\hspace{-0.5mm}\right)}}{\sqrt{2\pi np'(\hspace{-0.25mm}1\hspace{-0.5mm}-\hspace{-0.5mm}p'\hspace{-0.25mm})}},~~~~~~~~
\eea
using which it is straightforward to obtain from (\ref{procakant}) that
\bea \nonumber
z'' = z' - z' \frac{e^{-\frac{\mu^2}{2p'(1-p')n}+\mathcal{O}\left(\frac{\mu}{n}\right)}}{\sqrt{2\pi np'(1-p')}}\left(1-e^{-\beta'  (n z' + 2\mu)E}\right).
\eea
Clearly, for $\mu\leq\mathcal{O}(\sqrt{n})$, 
\bea \label{rezo}
z''=z'-\mathcal{O}(1/\sqrt{n}).
\eea

On the other hand, observe that the left hand side of (\ref{banadzev}) is the population of the level $np'-\mu$, and the summation of these values over all $\mu\leq\mathcal{O}(\sqrt{n})$, will produce $1-\mathcal{O}(1/\sqrt{n})$. Hence, if we apply the inversions described by $V$ on all levels with $\mu\leq\mathcal{O}(\sqrt{n})$, we will arrive at a state with local bias being $-z'+\mathcal{O}(1/\sqrt{n})$. Now, since each inversion changes the initial bias by $\mathcal{O}(1/\sqrt{n})$ (\ref{rezo}), we conclude that by conducting a sequence $\mathcal{O}(\sqrt{n})$ steps, one can change the initial local bias $z'$ to any $|z|<z'$, with the precision increasing with $n$. Therefore, in thermodynamic limit there exist diagonal states which asymptotically saturate the thermodynamic bound (\ref{wfixs}).

\section{Correlations in degenerate subspaces}\label{appc}

Consider the total Hamiltonian
\begin{equation}
H=\sum_{i=1}^{n}h_i= \sum_{i=1}^{n_{l}} E_i \Pi_i, 
\end{equation}
where each  $h_i=h:=\sum_{a=0}^{d-1}\epsilon_a |a\rangle\langle a|$ (with $\epsilon_0=0$) has local dimension $d$, which we assume to be finite. The number of different global energies, $n_l$ is found to be
\begin{equation}
n_l=C_{n+d-1}^{d-1}=\frac{(n+d-1)!}{n! (d-1)!},
\end{equation}
which corresponds to the number of non-zero eigenvalues of (\ref{CorrDeg}). In order to find the passive state associated to (\ref{CorrDeg}), one has to move such eigenvalues to the lowest energy levels. This operation requires knowledge of the spectrum of $h_i$. Nevertheless, it will suffice for our purposes to move them to a sufficiently degenerated energy. The degeneracy of a global energy $E_i=\sum_j k^{(i)}_j \epsilon_j$ is equal to  $C_{n}^{k^{(i)}_1,k^{(i)}_2,...,k^{(i)}_d}$. The point is then to find the lowest energy, $E_{\rm min}$, satisfying $C_{n}^{k^{\rm min}_1,k^{\rm min}_2,...,k^{\rm min}_d}\geq C_{n+d-1}^{d-1}$, so that the work extracted after such a transformation is simply given by
\bea
W_{\rm deg}\geq E_{\rho_{\deg}}-E_{\rm min}.
\eea
Now, notice that for large $n$
\bea
\lim_{n\rightarrow \infty} \frac{C_{n+d-1}^{d-1}}{C_{n}^{n-d,k^{'}_2,...,k^{'}_d}}=0, \hspace{7mm} \sum_{j=2}^d k^{'}_j =d
\eea
with $E'=\sum_{a=2}^d k^{'}_a \epsilon_a$. Observe that $E'$ is of the order of the energy of one subsystem (for instance, choosing $k'_2=d$ and $k'_j=0$ for $j>2$, we obtain $E'=d \epsilon_2$). Therefore we can take $E_{\rm min}=E'$ obtaining the desired result.

In the case of $d=2$ the expression for $E_{\rm min}$ is particularly simple:
\bea
W_{\rm min}=\left[1- C_n^{pn} p^{np} (1-p)^{(1-p)n} \right] E.
\eea


\begin{thebibliography}{99}
\bibitem{commentsp} An even stronger notion of local passivity was recently introduced in \cite{funo}, where local states from which no work can be extracted by any quantum operation (i.e., any Positive-Valued-Operator-Measure (POVM)) were studied. For that to happen the global state must be entangled \cite{funo}.
\bibitem{zerotemp} This is true unless $T=0$ (or $\beta=\infty$). On the other hand, in that case $\tau_\infty=\pr{0}$ and $E_\infty=0$; i.e., there is no work to extract and no correlations to employ.
\bibitem{fuflfoot} For qubits or qutrits ($d=$ 2 and 3, respectively) any ensemble is big enough, since in both cases $d-1\leq 2$.
\bibitem{yishk} The solution is unique because the entropy of the thermal state, $S(\tau_\beta)$, is a monotonic function of $\beta$.
\bibitem{ykh} Here we provide a detailed analysis for the case of qubits. For the general case of qudits ($d$-level systems) see \cite{us}, where the same protocol is applied for creating correlations at minimal energy cost.
\bibitem{footx} The inequality is strict so that, e.g., when $n$ is even, the states with $|\iv|=|\ivb|=n/2$ should remain untouched.
\bibitem{toof} That is, the system needs to undergo an arbitrarily large number of infinitesimal collisions with the bath, so that during each interaction the back action is negligible.
\bibitem{hajat} Note that to saturate (\ref{fed}), the bath has to be infinitely large.
\end{thebibliography}
\end{document}